\documentclass[prd,aps,showpacs,epsf,floats,10pt]{revtex4}
\usepackage{amssymb}
\usepackage{amsfonts}
\usepackage{amsmath}

\input{tcilatex}
\begin{document}

\title{On The Complexity Of Statistical Models Admitting Correlations}
\author{Carlo Cafaro}
\email{carlo.cafaro@unicam.it}
\affiliation{Dipartimento di Fisica, Universit\`{a} di Camerino, I-62032 Camerino, Italy}
\author{Stefano Mancini}
\email{stefano.mancini@unicam.it}
\affiliation{Dipartimento di Fisica, Universit\`{a} di Camerino, I-62032 Camerino, Italy}

\begin{abstract}
We compute the asymptotic temporal behavior of the dynamical complexity
associated with the maximum probability trajectories on Gaussian statistical
manifolds in presence of correlations between the variables labeling the
macrostates of the system. The algorithmic structure of our asymptotic
computations is presented and special focus is devoted to the
diagonalization procedure that allows to simplify the problem in a
remarkable way. We observe a power law decay of the information geometric
complexity at a rate determined by the correlation coefficient. We conclude
that macro-correlations lead to the emergence of an asymptotic \emph{%
information geometric compression} of the statistical macrostates explored
on the configuration manifold of the model in its evolution between the
initial and final macrostates.
\end{abstract}

\pacs{%
Probability
Theory
(02.50.Cw),
Riemannian
Geometry
(02.40.Ky),
Chaos
(05.45.-a),
Complexity (89.70.Eg),
Entropy
(89.70.Cf).%
}
\maketitle

\section{\textbf{Introduction}}

The study of complexity \cite{gell-mann} has created a new set of ideas on
how very simple systems may give rise to very complex behaviors. Chaotic
behavior is a particular case of complex behavior and it will be the object
of the present work. In this paper we make use of the so-called Entropic
Dynamics (ED) \cite{caticha1} and Information Geometrodynamical Approach to
Chaos (IGAC) \cite{carlo-tesi, carlo-CSF}. ED is a theoretical framework
that arises from the combination of inductive inference (Maximum Entropy
Methods, \cite{caticha2}) and Information Geometry \cite{amari}. The most
intriguing question being pursued in ED stems from the possibility of
deriving dynamics from purely entropic arguments. This is clearly valuable
in circumstances where microscopic dynamics may be too far removed from the
phenomena of interest, such as in complex biological or ecological systems,
or where it may just be unknown or perhaps even nonexistent, as in
economics. The applicability of ED has been extended to temporally-complex
(chaotic) dynamical systems on curved statistical manifolds and relevant
measures of chaoticity of such an IGAC have been identified \cite{carlo-tesi}%
. IGAC arises as a theoretical framework to study chaos in informational
geodesic flows describing physical, biological or chemical \ systems. A
geodesic on a curved statistical manifold $\mathcal{M}_{S}$ represents the
maximum probability path a complex dynamical system explores in its
evolution between initial and final macrostates. Each point of the geodesic
is parametrized by the macroscopic dynamical variables\textbf{\ }$\left\{
\Theta \right\} $ defining the macrostate of the system. Furthermore, each
macrostate is in a one-to-one correspondence with the probability
distribution $p\left( X|\Theta \right) $ representing the maximally probable
description of the system being considered. The set of macrostates forms the
parameter space $\mathcal{D}_{\Theta }$ while the set of probability
distributions forms the statistical manifold $\mathcal{M}_{S}$. We point out
that the knowledge of both initial and final states is not necessary for
carrying out macroscopic predictions. For instance, another possible
situation may be the one where we only know the initial state and assume
that the system evolves to other states. In such a case, it can be shown
that the system moves irreversibly and continuously along the entropy
gradient \cite{catichaIED}.

IGAC\ is the information geometric analogue of conventional
geometrodynamical approaches \cite{casetti, di bari} where the classical
configuration space $\Gamma _{E}$\ is replaced by a statistical manifold $%
\mathcal{M}_{S}$\ with the additional possibility of considering chaotic
dynamics arising from non conformally flat metrics (the Jacobi metric is
always conformally flat, instead). It is an information geometric extension
of the Jacobi geometrodynamics (the geometrization of a Hamiltonian system
by transforming it to a geodesic flow \cite{jacobi}). The reformulation of
dynamics in terms of a geodesic problem allows the application of a wide
range of well-known geometrical techniques in the investigation of the
solution space and properties of the equation of motion. The power of the
Jacobi reformulation is that all of the dynamical information is collected
into a single geometric object in which all the available manifest
symmetries are retained- the manifold on which geodesic flow is induced. For
example, integrability of the system is connected with existence of Killing
vectors and tensors on this manifold. The sensitive dependence of
trajectories on initial conditions, which is a key ingredient of chaos, can
be investigated from the equation of geodesic deviation. In the Riemannian 
\cite{casetti} and Finslerian \cite{di bari} (a Finsler metric is obtained
from a Riemannian metric by relaxing the requirement that the metric be
quadratic on each tangent space) geometrodynamical approach to chaos in
classical Hamiltonian systems, an active field of research concerns the
possibility of finding a rigorous relation among the sectional curvature,
the Lyapunov exponents, and the Kolmogorov-Sinai dynamical entropy (i. e.
the sum of positive Lyapunov exponents) \cite{kawabe}.

In this article, using statistical inference and information geometric
techniques, we investigate the macroscopic behavior of suitable complex
systems in terms of the underlying statistical structure of its microscopic
degrees of freedom in the presence of correlations. We compute the
asymptotic temporal behavior of the dynamical complexity of the maximum
probability trajectories on Gaussian statistical manifolds in presence of
correlation-like terms between the macrovariables labeling the macrostates
of the system. We observe a power law decay of the information geometric
complexity at a rate determined by the correlation coefficient.
Macro-correlations lead to the emergence of an asymptotic information
geometric compression of the explored statistical macro-states on the
configuration manifold of the model in its evolution between the initial and
final macrostates.

The layout of the paper is as follows. In the next Section, we briefly
discuss some differences between statistical models in absence and presence
of correlations. In Section III, we introduce the Gaussian statistical
manifold in presence of non-trivial off-diagonal terms. We find the Ricci
scalar curvature and the geodesic equations for such a system. In Section
IV, we introduce the information geometric diagonalization procedure that
allows to simplify the integration of the coupled set of nonlinear ordinary
differential equations leading to the maximally probable trajectories of our
model. In Section V, we compute the asymptotic temporal behavior of the
dynamical complexity of the geodesic trajectories for the correlated
two-dimensional Gaussian statistical model. In Section VI, we outline the
main steps involved in computing the asymptotic temporal behavior of the
dynamical complexity of geodesic trajectories for the $2l$-dimensional
macro-correlated Gaussian statistical model. Finally, in Section VI we
present final remarks and suggest further research directions.

\section{Statistical Models and Correlations}

In this Section, we introduce the notion of statistical models (manifolds)
in absence and presence of correlations between the microscopic degrees of
freedom of the system (micro-correlations) on the one hand and between the
variables \ labelling the macrostates of the system (macro-correlations) on
the other. For the sake of simplicity, we focus our attention on Gaussian
statistical models.

\subsection{Gaussian Statistical Models without Correlations}

Consider a statistical model whose microstates span a $l$-dimensional space
labelled by the variables $\left\{ X\right\} =\left\{ x_{1}\text{, }x_{2}%
\text{,..., }x_{l}\right\} $ with $x_{j}\in 
%TCIMACRO{\U{211d} }%
%BeginExpansion
\mathbb{R}
%EndExpansion
$, $\forall j=1$,..., $l$. We assume the only testable information
pertaining to the quantities $x_{j}$ consists of the expectation values $%
\left\langle x_{j}\right\rangle $ and the variance $\Delta x_{j}$. This set
of expected values define the $2l$-dimensional space of macrostates of the
system. A measure of distinguishability among the macrostates of the model
is achieved by assigning a probability distribution $P\left( X|\Theta
\right) $ to each $2l$-dimensional macrostate $\Theta \overset{\text{def}}{=}%
\left\{ \left( ^{\left( 1\right) }\theta _{j}\text{,}^{\left( 2\right)
}\theta _{j}\right) \right\} _{l\text{-pairs}}$ $=\left\{ \left(
\left\langle x_{j}\right\rangle \text{, }\Delta x_{j}\right) \right\} _{l%
\text{-pairs}}$. The process of assigning a probability distribution to each
state provides $\mathcal{M}_{S}$ with a metric structure. Specifically, the
Fisher-Rao information metric $g_{\mu \nu }\left( \Theta \right) $ \cite%
{amari} is a measure of distinguishability among macrostates on the
statistical manifold $\mathcal{M}_{S}$ \textbf{(}see (\ref{manifold}\textbf{%
))}, 
\begin{equation}
g_{\mu \nu }\left( \Theta \right) =\int dXP\left( X|\Theta \right) \partial
_{\mu }\log P\left( X|\Theta \right) \partial _{\nu }\log P\left( X|\Theta
\right) =4\int dX\partial _{\mu }\sqrt{P\left( X|\Theta \right) }\partial
_{\nu }\sqrt{P\left( X|\Theta \right) }\text{,}  \label{FRM}
\end{equation}%
with $\mu $, $\nu =1$,..., $2l$ and $\partial _{\mu }=\frac{\partial }{%
\partial \Theta ^{\mu }}$. It assigns an information geometry to the space
of states. The information metric $g_{\mu \nu }\left( \Theta \right) $ is a
symmetric and positive definite Riemannian metric. For the sake of
completeness and in view of its potential relevance in the study of
correlations, we point out that the Fisher-Rao metric satisfies the
following two properties: 1) invariance under (invertible) transformations
of microvariables $\left\{ x\right\} \in \mathcal{X}$\textbf{;} 2)
covariance under reparametrization of the statistical macrospace $\left\{
\theta \right\} \in \mathcal{D}_{\theta }$. The invariance of $g_{\mu \nu
}\left( \theta \right) $ under reparametrization of the microspace $\mathcal{%
X}$ implies that \cite{amari}\textbf{,}%
\begin{equation}
\mathcal{X}\subseteq 
%TCIMACRO{\U{211d} }%
%BeginExpansion
\mathbb{R}
%EndExpansion
^{l}\ni x\longmapsto y\overset{\text{def}}{=}f\left( x\right) \in \mathcal{Y}%
\subseteq 
%TCIMACRO{\U{211d} }%
%BeginExpansion
\mathbb{R}
%EndExpansion
^{l}\Longrightarrow p\left( x|\theta \right) \longmapsto p^{\prime }\left(
y|\theta \right) =\left[ \frac{1}{\left\vert \frac{\partial f}{\partial x}%
\right\vert }p\left( x|\theta \right) \right] _{x=f^{-1}\left( y\right) }%
\text{.}
\end{equation}%
The covariance under reparametrization of the parameter space\textbf{\ }$%
\mathcal{D}_{\theta }$\textbf{\ }(homeomorphic to\textbf{\ }$\mathcal{M}_{S}$%
\textbf{) }implies that \cite{amari},%
\begin{equation}
\mathcal{D}_{\theta }\ni \theta \longmapsto \theta ^{\prime }\overset{\text{%
def}}{=}f\left( \theta \right) \in \mathcal{D}_{\theta ^{\prime
}}\Longrightarrow g_{\mu \nu }\left( \theta \right) \longmapsto g_{\mu \nu
}^{\prime }\left( \theta ^{\prime }\right) =\left[ \frac{\partial \theta
^{\alpha }}{\partial \theta ^{\prime \mu }}\frac{\partial \theta ^{\beta }}{%
\partial \theta ^{\prime \nu }}g_{\alpha \beta }\left( \theta \right) \right]
_{\theta =f^{-1}\left( \theta ^{\prime }\right) }\text{,}
\end{equation}%
where,%
\begin{equation}
g_{\mu \nu }\left( \theta \right) =\int dxp\left( x|\theta \right) \partial
_{\mu }\log p\left( x|\theta \right) \partial _{\nu }\log p\left( x|\theta
\right) \text{ and, }g_{\mu \nu }^{\prime }\left( \theta ^{\prime }\right)
=\int dxp^{\prime }\left( x|\theta ^{\prime }\right) \partial _{\mu
}^{\prime }\log p^{\prime }\left( x|\theta ^{\prime }\right) \partial _{\nu
}\log p^{\prime }\left( x|\theta ^{\prime }\right) \text{,}
\end{equation}%
with $\partial _{\mu }^{\prime }=\frac{\partial }{\partial \theta ^{\prime
\mu }}$ and $p^{\prime }\left( x|\theta ^{\prime }\right) =p\left( x|\theta
=f^{-1}\left( \theta ^{\prime }\right) \right) $. Our $2l$\textbf{-}%
dimensional statistical model represents a macroscopic (probabilistic)
description of a microscopic $l$-dimensional (microscopic) physical system
evolving over\textbf{\ }a $l$-dimensional (micro) space.\ The variables $%
\left\{ X\right\} =\left\{ x_{1}\text{, }x_{2}\text{,..., }x_{l}\right\} $
label the $l$-dimensional space of microstates of the system. We assume that
all information relevant to the dynamical evolution of the system is
contained in the probability distributions. For this reason, no other
information is required. Each macrostate may be viewed\textbf{\ }as a point
of a $2l$-dimensional statistical manifold with coordinates given by the
numerical values of the expectations $^{\left( 1\right) }\theta _{j}$ and $%
^{\left( 2\right) }\theta _{j}$. The available \textit{relevant information}
can be written in the form of the following $2l$ information constraint
equations,%
\begin{equation}
\left\langle x_{j}\right\rangle =\dint\limits_{-\infty }^{+\infty
}dx_{j}x_{j}P_{j}\left( x_{j}|^{\left( 1\right) }\theta _{j}\text{,}^{\left(
2\right) }\theta _{j}\right) \text{, }\Delta x_{j}=\left[ \dint\limits_{-%
\infty }^{+\infty }dx_{j}\left( x_{j}-\left\langle x_{j}\right\rangle
\right) ^{2}P_{j}\left( x_{j}|^{\left( 1\right) }\theta _{j}\text{,}^{\left(
2\right) }\theta _{j}\right) \right] ^{\frac{1}{2}}\text{.}  \label{C1}
\end{equation}%
The probability distributions $P_{j}$ in (\ref{C1}) are constrained by the
conditions of normalization,%
\begin{equation}
\dint\limits_{-\infty }^{+\infty }dx_{j}P_{j}\left( x_{j}|^{\left( 1\right)
}\theta _{j}\text{,}^{\left( 2\right) }\theta _{j}\right) =1\text{.}
\label{C2}
\end{equation}%
Information theory identifies the Gaussian distribution as the maximum
entropy distribution if only the expectation value and the variance are
known \cite{tribus}\textbf{.} Maximum relative Entropy methods \cite%
{caticha(REII), caticha-giffin, caticha2} allow us to associate a
probability distribution $P\left( X|\Theta \right) $ to each point in the
space of states $\Theta $. The distribution that best reflects the
information contained in the prior distribution $m\left( X\right) $ updated
by the information $\left( \left\langle x_{j}\right\rangle \text{, }\Delta
x_{j}\right) $ is obtained by maximizing the relative entropy, 
\begin{equation}
S\left( \Theta \right) =-\int d^{l}XP\left( X|\Theta \right) \log \left( 
\frac{P\left( X|\Theta \right) }{m\left( X\right) }\right) \text{,}
\label{RE}
\end{equation}%
where $m\left( X\right) $ is the prior probability distribution. As a
working hypothesis, the prior $m\left( X\right) $ is set to be uniform since
we assume the lack of prior available information about the system \cite%
{newJay}. We assume uncoupled constraints among microvariables\textbf{\ }$%
x_{j}$\textbf{.} In other words, we assume that information about
correlations between the microvariables need not to be tracked. Therefore,
upon maximizing (\ref{RE}) given the constraints (\ref{C1}) and (\ref{C2}),
we obtain%
\begin{equation}
P\left( X|\Theta \right) =\dprod\limits_{j=1}^{l}P_{j}\left( x_{j}|^{\left(
1\right) }\theta _{j}\text{,}^{\left( 2\right) }\theta _{j}\right)
\label{PDG}
\end{equation}%
where%
\begin{equation}
P_{j}\left( x_{j}|^{\left( 1\right) }\theta _{j}\text{,}^{\left( 2\right)
}\theta _{j}\right) =\left( 2\pi \sigma _{j}^{2}\right) ^{-\frac{1}{2}}\exp %
\left[ -\frac{\left( x_{j}-\mu _{j}\right) ^{2}}{2\sigma _{j}^{2}}\right]
\end{equation}%
and,\textbf{\ }in standard notation for Gaussians, $^{\left( 1\right)
}\theta _{^{j}}\overset{\text{def}}{=}\left\langle x_{j}\right\rangle \equiv
\mu _{j}$, $^{\left( 2\right) }\theta _{j}\overset{\text{def}}{=}\Delta
x_{j}\equiv \sigma _{j}$. The probability distribution (\ref{PDG}) encodes
the available information concerning the system. The statistical manifold $%
\mathcal{M}_{S}$ associated to (\ref{PDG}) is formally defined as follows,%
\begin{equation}
\mathcal{M}_{S}=\left\{ P\left( X|\Theta \right) =\underset{j=1}{\overset{l}{%
\dprod }}P_{j}\left( x_{j}|\mu _{j}\text{, }\sigma _{j}\right) \right\} 
\text{,}  \label{manifold}
\end{equation}%
where $X\in 
%TCIMACRO{\U{211d} }%
%BeginExpansion
\mathbb{R}
%EndExpansion
^{l}$ and $\Theta $ belongs to the $2l$-dimensional parameter space $%
\mathcal{D}_{\Theta }=\left[ \mathcal{I}_{\mu }\times \mathcal{I}_{\sigma }%
\right] ^{l}$\textbf{. }The parameter space $\mathcal{D}_{\Theta }$
(homeomorphic to\textbf{\ }$\mathcal{M}_{S}$\textbf{)} is the direct product
of the parameter subspaces $\mathcal{I}_{\mu }$ and $\mathcal{I}_{\sigma }$%
\textbf{, }where (in the Gaussian case, unless specified otherwise) $%
\mathcal{I}_{\mu }=\left( -\infty \text{, }+\infty \right) _{\mu }$ and $%
\mathcal{I}_{\sigma }=\left( 0\text{, }+\infty \right) _{\sigma }$\textbf{. }%
The line element $ds^{2}=g_{\mu \nu }\left( \Theta \right) d\Theta ^{\mu
}d\Theta ^{\nu }$ arising from (\ref{PDG}) is \cite{cafaroIJTP},%
\begin{equation}
ds_{\mathcal{M}_{s}}^{2}\overset{\text{def}}{=}\dsum\limits_{j=1}^{l}\left( 
\frac{1}{\sigma _{j}^{2}}d\mu _{j}^{2}+\frac{2}{\sigma _{j}^{2}}d\sigma
_{j}^{2}\right) \text{, with }\mu \text{, }\nu =1\text{,..., }2l\text{.}
\end{equation}

\subsection{Gaussian Statistical Models with Correlations}

Coupled constraints would lead to a "generalized" product rule in (\ref{PDG}%
) and to a metric tensor (\ref{FRM}) with \textit{non-trivial off-diagonal
elements} (covariance terms). In presence of micro-correlated degrees of
freedom $\left\{ x_{j}\right\} $, the "generalized" product rule becomes,%
\begin{equation}
P_{\text{tot}}\left( x_{1}\text{,..., }x_{l}\right)
=\dprod\limits_{j=1}^{l}P_{j}\left( x_{j}\right) \overset{\text{correlations}%
}{\longrightarrow }P_{\text{tot}}^{\prime }\left( x_{1}\text{,..., }%
x_{l}\right) \neq \dprod\limits_{j=1}^{l}P_{j}\left( x_{j}\right) \text{,}
\end{equation}%
where,%
\begin{equation}
P_{\text{tot}}^{\prime }\left( x_{1}\text{,..., }x_{l}\right) \overset{\text{%
def}}{=}P_{l}\left( x_{l}|x_{1}\text{,..., }x_{l-1}\right) P_{l-1}\left(
x_{l-1}|x_{1}\text{,..., }x_{l-2}\right) \text{...}P_{2}\left(
x_{2}|x_{1}\right) P_{1}\left( x_{1}\right) \text{.}  \label{pp}
\end{equation}%
Correlations in the microscopic degrees of freedom may be introduced in
terms of the following information-constraints,%
\begin{equation}
x_{j}=f_{j}\left( x_{1}\text{,..., }x_{j-1}\right) \text{, }\forall j=2\text{%
,..., }l\text{.}
\end{equation}%
In such a case (\ref{pp}) becomes%
\begin{equation}
P_{\text{tot}}^{\prime }\left( x_{1}\text{,..., }x_{l}\right) =\delta \left(
x_{l}-f_{l}\left( x_{1}\text{,..., }x_{l-1}\right) \right) \delta \left(
x_{l-1}-f_{l-1}\left( x_{1}\text{,..., }x_{l-2}\right) \right) ...\delta
\left( x_{2}-f_{2}\left( x_{1}\right) \right) P_{1}\left( x_{1}\right) \text{%
,}
\end{equation}%
where the $j$-th probability distribution $P_{j}\left( x_{j}\right) $ is
given by,%
\begin{equation}
P_{j}\left( x_{j}\right) =\int \text{...}\int dx_{1}\text{...}%
dx_{j-1}dx_{j+1}\text{...}dx_{l}P_{\text{tot}}^{\prime }\left( x_{1}\text{%
,..., }x_{l}\right) \text{.}
\end{equation}%
Correlations between the microscopic degrees of freedom of the system $%
\left\{ x_{j}\right\} $ (\textit{microcorrelations}) are conventionally
introduced by means of the correlation coefficients $r_{ij}^{\left( \text{%
micro}\right) }$, 
\begin{equation}
r_{ij}^{\left( \text{micro}\right) }=r\left( x_{i}\text{, }x_{j}\right) 
\overset{\text{def}}{=}\frac{\left\langle x_{i}x_{j}\right\rangle
-\left\langle x_{i}\right\rangle \left\langle x_{j}\right\rangle }{\sigma
_{i}\sigma _{j}}\text{, with }\sigma _{i}=\sqrt{\left\langle \left(
x_{i}-\left\langle x_{i}\right\rangle \right) ^{2}\right\rangle }\text{,}
\end{equation}%
with $r_{ij}^{\left( \text{micro}\right) }\in \left( -1\text{, }1\right) $
and $i$, $j=1$,..., $l$. For the present $2l$-dimensional Gaussian
statistical model in presence of microcorrelations, the system is described
by the following probability distribution $P\left( X|\Theta \right) $,%
\begin{equation}
P\left( X|\Theta \right) =\frac{1}{\left[ \left( 2\pi \right) ^{l}\det
C\left( \Theta \right) \right] ^{\frac{1}{2}}}\exp \left[ -\frac{1}{2}\left(
X-M\right) ^{t}\cdot C^{-1}\left( \Theta \right) \cdot \left( X-M\right) %
\right] \neq \dprod\limits_{j=1}^{l}\left( 2\pi \sigma _{j}^{2}\right) ^{-%
\frac{1}{2}}\exp \left[ -\frac{\left( x_{j}-\mu _{j}\right) ^{2}}{2\sigma
_{j}^{2}}\right] \text{,}  \label{CG}
\end{equation}%
where $X=\left( x_{1}\text{,..., }x_{l}\right) $, $M=\left( \mu _{1}\text{%
,..., }\mu _{l}\right) $ and $C\left( \Theta \right) $ is the $\left(
2l\times 2l\right) $-dimensional (non-singular) covariance matrix.

As a first step in studying the introduction of correlations into
statistical models, we will focus on the relatively simple (but non trivial)
model introduced in the following paragraph. In principle, the computational
techniques introduced in this article would be valid for arbitrary complex
statistical models in presence of correlations. Additional difficulties
arise purely from a computational point of view.

\section{The Two-Dimensional Macro-correlated Gaussian Statistical Model}

In this Section we study the \ information dynamics of a system with $l$
degrees of freedom, each described by two \textit{correlated} pieces of
relevant information, its mean expected value and its variance (Gaussian
statistical macrostates in presence of macro-correlations). The line element 
$ds^{2}=g_{\mu \nu }\left( \Theta \right) d\Theta ^{\mu }d\Theta ^{\nu }$ on 
$\mathcal{M}_{s}$ that we consider is defined by,%
\begin{equation}
ds_{\mathcal{M}_{s}}^{2}\overset{\text{def}}{=}\sum_{j=1}^{l}\left( \frac{1}{%
\sigma _{j}^{2}}d\mu _{j}^{2}+\frac{2r_{j}}{\sigma _{j}^{2}}d\mu _{j}d\sigma
_{j}+\frac{2}{\sigma _{j}^{2}}d\sigma _{j}^{2}\right) \text{, with }\mu 
\text{, }\nu =1\text{,..., }2l\text{.}
\end{equation}%
We restrict our consideration to positive coefficients $r_{j}\in \left( 0%
\text{, }1\right) $, $\forall j=1$,..., $l$ and we assume they play the role
of macro-correlations between the macro-variables of the system. This leads
to consider a $2l$-dimensional statistical manifold $\mathcal{M}_{s}$ where
the probability distributions representing the state of the system exhibit
deviations from standard Gaussian probability distributions due to the
introduction of nonvanishing coefficients $r_{k}$. We believe the study of
this relatively simplified model constitutes an important preliminary step
towards the computation of the asymptotic behavior of the dynamical
complexity of microscopically correlated multidimensional Gaussian
statistical models. In particular, it improves our understanding of the
effect of correlations in the behavior of the complexity of maximally
probable trajectories of statistical models used to characterize systems in
a probabilistic way. For the sake of clarity, we will first consider the
two-dimensional Gaussian model with non-trivial off-diagonal terms.\textbf{\ 
}We then generalize our formalism in a straightforward way to the case of a%
\textbf{\ }$2l$\textbf{-}dimensional statistical manifold in Section VI.

\subsection{Information Geometry of the two-dimensional Model}

Consider the \ information dynamics of a system with $l=1$ microscopic
degree of freedom described by two \textit{correlated} pieces of relevant
information, its mean expected value and its variance. The Fisher-Rao line
element $ds^{2}$ of such a statistical model $\mathcal{M}_{s}$ is given by,%
\begin{equation}
ds_{\mathcal{M}_{s}}^{2}\overset{\text{def}}{=}g_{ij}\left( \theta \right)
d\theta ^{i}d\theta ^{j}=\frac{1}{\sigma ^{2}}d\mu ^{2}+\frac{2r}{\sigma ^{2}%
}d\mu d\sigma +\frac{2}{\sigma ^{2}}d\sigma ^{2}\text{, with }i\text{, }j=1%
\text{, }2\text{.}  \label{sm}
\end{equation}%
The Fisher-Rao information tensor $g_{ij}\left( \mu \text{, }\sigma \right) $
leading to the line element in (\ref{sm}) is given by,%
\begin{equation}
g_{ij}\left( \mu \text{, }\sigma \right) =\frac{1}{\sigma ^{2}}\left( 
\begin{array}{cc}
1 & r \\ 
r & 2%
\end{array}%
\right) \text{.}  \label{a}
\end{equation}%
Therefore, the inverse metric tensor $g^{ij}\left( \mu \text{, }\sigma
\right) $ becomes, 
\begin{equation}
g^{ij}\left( \mu \text{, }\sigma \right) =\frac{\sigma ^{2}}{2-r^{2}}\left( 
\begin{array}{cc}
2 & -r \\ 
-r & 1%
\end{array}%
\right) \text{.}  \label{b}
\end{equation}%
The metric tensor $g_{ij}\left( \mu \text{, }\sigma \right) $ and its
inverse $g^{ij}\left( \mu \text{, }\sigma \right) $ are useful quantities
for the computation of the Christoffel connection coefficients $\Gamma
_{ij}^{k}$ of the curved statistical manifold (model) $\mathcal{M}_{s}$.
Recall that the connection coefficients $\Gamma _{ij}^{k}$ are defined as 
\cite{felice},%
\begin{equation}
\Gamma _{ij}^{k}\overset{\text{def}}{=}\frac{1}{2}g^{km}\left( \partial
_{i}g_{mj}+\partial _{j}g_{im}-\partial _{m}g_{ij}\right) \text{.}  \label{c}
\end{equation}%
In our case, we have to compute the seven non-vanishing connection
coefficients $\Gamma _{11}^{1}$, $\Gamma _{11}^{2}$, $\Gamma
_{12}^{2}=\Gamma _{21}^{2}$, $\Gamma _{22}^{2}$, $\Gamma _{11}^{1}$, $\Gamma
_{12}^{1}=\Gamma _{21}^{1}$, $\Gamma _{22}^{1}$. Substituting (\ref{a}) and (%
\ref{b}) in (\ref{c}), we obtain,%
\begin{eqnarray}
\Gamma _{11}^{1} &=&-\frac{r}{2-r^{2}}\frac{1}{\sigma }\text{, }\Gamma
_{11}^{2}=\frac{1}{2-r^{2}}\frac{1}{\sigma }\text{, }\Gamma _{21}^{1}=\Gamma
_{12}^{1}=-\frac{2}{2-r^{2}}\frac{1}{\sigma }\text{, }  \notag \\
&&  \notag \\
\Gamma _{12}^{2} &=&\Gamma _{21}^{2}=\frac{r}{2-r^{2}}\frac{1}{\sigma }\text{%
, }\Gamma _{22}^{1}=-\frac{2r}{2-r^{2}}\frac{1}{\sigma }\text{, }\Gamma
_{22}^{2}=\frac{2r^{2}-2}{2-r^{2}}\frac{1}{\sigma }\text{.}  \label{A}
\end{eqnarray}%
Notice that in the limit $r\rightarrow 0$, we get the Christoffel connection
coefficients for the two-dimensional Gaussian statistical model in absence
off-diagonal matrix elements in $g_{ij}\left( \mu \text{, }\sigma \right) $ 
\cite{cafaroPD}. Having obtained the non-vanishing $\Gamma _{ij}^{k}$, we
compute the Ricci curvature tensor $\mathcal{R}_{ij}$ defined as \cite%
{felice},%
\begin{equation}
\mathcal{R}_{ij}\overset{\text{def}}{=}\partial _{k}\Gamma
_{ij}^{k}-\partial _{j}\Gamma _{ik}^{k}+\Gamma _{ij}^{k}\Gamma
_{kn}^{n}-\Gamma _{ik}^{m}\Gamma _{jm}^{k}\text{.}  \label{e}
\end{equation}%
Substituting (\ref{A}) in (\ref{e}), we obtain the three non-vanishing $%
\mathcal{R}_{ij}$ components,%
\begin{equation}
\mathcal{R}_{11}=-\frac{1}{2-r^{2}}\frac{1}{\sigma ^{2}}\text{, }\mathcal{R}%
_{12}=\mathcal{R}_{21}=-\frac{r}{2-r^{2}}\frac{1}{\sigma ^{2}}\text{, }%
\mathcal{R}_{22}=-\frac{2}{2-r^{2}}\frac{1}{\sigma ^{2}}\text{.}  \label{f}
\end{equation}%
Again, notice that in the limit $r\rightarrow 0$, we get the standard Ricci
tensor components in absence of macro-correlations in the Gaussian model.
Finally, we compute Ricci scalar curvature $\mathcal{R}_{\mathcal{M}_{s}}$,%
\begin{equation}
\mathcal{R}_{\mathcal{M}_{s}}\overset{\text{def}}{=}\mathcal{R}_{ij}g^{ij}%
\text{.}  \label{g}
\end{equation}%
Substituting (\ref{b}) and (\ref{f}) in (\ref{g}), $\mathcal{R}_{\mathcal{M}%
_{s}}\left( r\right) $ becomes,%
\begin{equation}
\mathcal{R}_{\mathcal{M}_{s}}\left( r\right) =g^{11}\mathcal{R}_{11}+2g^{12}%
\mathcal{R}_{12}+g^{22}\mathcal{R}_{22}=-\frac{2}{2-r^{2}}\text{.}
\end{equation}%
In absence of correlations, $r\rightarrow 0$, we recover the known relation $%
\mathcal{R}_{\mathcal{M}_{s}}=-1$.

\subsection{Information Dynamics on $\mathcal{M}_{s}$}

The information dynamics can be derived from a standard principle of least
action of Jacobi type \cite{caticha1}. The geodesic equations for the
macrovariables of the Gaussian ED model are given by\textit{\ nonlinear}
second order coupled ordinary differential equations,%
\begin{equation}
\frac{d^{2}\Theta ^{\mu }}{d\tau ^{2}}+\Gamma _{\nu \rho }^{\mu }\frac{%
d\Theta ^{\nu }}{d\tau }\frac{d\Theta ^{\rho }}{d\tau }=0\text{.}  \label{GE}
\end{equation}%
The geodesic equations in (\ref{GE}) describe a \textit{reversible} dynamics
whose solution is the trajectory between an initial $\Theta ^{\left( \text{%
initial}\right) }$ and a final macrostate $\Theta ^{\left( \text{final}%
\right) }$. The trajectory can be equally well traversed in both directions 
\cite{caticha1}.

As mentioned in the Introduction, there are two different scenarios explored
within the framework of entropic dynamics. In the first scenario (\emph{%
irreversible entropic dynamics}, \cite{catichaIED}) , assuming that the
system evolves from a given initial state to other states, the trajectory
that the system is expected to follow can be studied. In this problem, the
existence of a trajectory is assumed and, in addition, it is assumed that
information about the initial state is sufficient to determine it. The
application of a principle of inference (the method of maximum entropy, ME)
to the only information available (the initial state) and the recognition
that motion occurred leads to the dynamical law. The resulting entropic
dynamics is very simple: the system moves irreversibly and continuously
along the entropy gradient. However, the question of whether the actual
trajectory is the expected one remains unanswered and it depends on whether
the information encoded in the initial state happened to be sufficient for
prediction. For many systems more information is needed, even for those for
which the dynamics is reversible. In the reversible case (\emph{reversible
entropic dynamics}, \cite{caticha1}), assuming that the system evolves from
a given initial state to a given final state, the objective is to study what
trajectory is the system expected to follow. Again, it is implicitly assumed
that there is a trajectory, that in moving from one state to another the
system will pass through a continuous set of intermediate states. The
equation of motion follows from a principle of inference, the principle of
maximum entropy, and not from a principle of physics. In the resulting
entropic dynamics, the system moves along a geodesic in the space of states.
The information geometry of this space is curved and possibly quite
complicated (as the one studied in this manuscript).\textbf{\ }The key-point
is that the statistical model supplies its own notion of time and since the
irreversible macroscopic motion is not explained in terms of a reversible
microscopic motion there is no need to explain irreversibility, this
question does not really arise. Furthermore, we emphasize that while the
dynamical law characterizing the irreversible entropic dynamics is \emph{%
linear} (one boundary condition, affine asymmetry- irreversibility) in the
affine parameter $\tau $, the one associated to the reversible entropic
dynamics is \emph{quadratic} (two boundary conditions, affine
symmetry-reversibility).

In the case under consideration, substituting (\ref{A}) in (\ref{GE}) and
noticing that $\theta ^{1}=\mu $, $\theta ^{2}=\sigma $, the geodesic
equations become,%
\begin{eqnarray}
0 &=&\frac{d^{2}\mu }{d\tau ^{2}}-\frac{r}{2-r^{2}}\frac{1}{\sigma }\left( 
\frac{d\mu }{d\tau }\right) ^{2}-\frac{4}{2-r^{2}}\frac{1}{\sigma }\frac{%
d\mu }{d\tau }\frac{d\sigma }{d\tau }-\frac{2r}{2-r^{2}}\frac{1}{\sigma }%
\left( \frac{d\sigma }{d\tau }\right) ^{2}\text{,}  \notag \\
&&  \notag \\
0 &=&\frac{d^{2}\sigma }{d\tau ^{2}}+\frac{1}{2-r^{2}}\frac{1}{\sigma }%
\left( \frac{d\mu }{d\tau }\right) ^{2}+\frac{2r}{2-r^{2}}\frac{1}{\sigma }%
\frac{d\mu }{d\tau }\frac{d\sigma }{d\tau }+\frac{2r^{2}-2}{2-r^{2}}\frac{1}{%
\sigma }\left( \frac{d\sigma }{d\tau }\right) ^{2}\text{.}  \label{GE1}
\end{eqnarray}%
Integration of the coupled system of nonlinear second order ordinary
differential equations in (\ref{GE1}) is highly non trivial, although the
system can be solved in the case $r=0$ \cite{cafaroPD}. Therefore, in what
follows we will introduce an \textit{information geometric diagonalization
procedure} that allows us to tackle such problem in a more convenient way.

\section{Information Geometric Diagonalization Procedure}

In this Section, we introduce the information geometric diagonalization
procedure that allows to simplify the integration of (\ref{GE1}).

\subsection{Preliminaries}

Consider the $2\times 2$ matrix representation of an arbitrary Fisher-Rao
information metric tensor $\hat{g}$ in a basis $\mathcal{B}_{\text{old}}$,%
\begin{equation}
\left[ \hat{g}\right] _{\mathcal{B}_{\text{old}}}=\left( 
\begin{array}{cc}
g\left( \partial _{1}\text{, }\partial _{1}\right) & g\left( \partial _{1}%
\text{, }\partial _{2}\right) \\ 
g\left( \partial _{2}\text{, }\partial _{1}\right) & g\left( \partial _{2}%
\text{, }\partial _{2}\right)%
\end{array}%
\right) \text{, }
\end{equation}%
The set $\mathcal{B}_{\text{old}}\overset{\text{def}}{=}\left\{ \partial _{1}%
\text{, }\partial _{2}\right\} $ with (in our case) $\partial _{1}=\frac{%
\partial }{\partial \mu }\equiv \partial _{\mu }$ and $\partial _{2}=\frac{%
\partial }{\partial \sigma }\equiv \partial _{\sigma }$ is a basis for the
tangent space at $p\in T_{p}\left( \mathcal{M}_{s}\right) $, where $\mathcal{%
M}_{s}$ is an arbitrary two-dimensional curved statistical manifold. An
arbitrary vector $\Theta $ in the tangent space $T_{p}\left( \mathcal{M}%
_{s}\right) $ can be written as,%
\begin{equation}
\Theta =\mu \partial _{\mu }+\sigma \partial _{\sigma }\text{.}
\end{equation}%
Recall that all non-singular matrices induce a basis transformation in $%
T_{p}\left( \mathcal{M}_{s}\right) $. For instance, consider the following
change of basis,%
\begin{equation}
\mathcal{B}_{\text{old}}\overset{\text{def}}{=}\left\{ \partial _{1}\text{, }%
\partial _{2}\right\} \longrightarrow \mathcal{B}_{\text{new}}\overset{\text{%
def}}{=}\left\{ \tilde{\partial}_{1}\text{, }\tilde{\partial}_{2}\right\} 
\text{,}
\end{equation}%
with $\tilde{\partial}_{1}=\frac{\partial }{\partial \tilde{\mu}}$ and $%
\tilde{\partial}_{2}=\frac{\partial }{\partial \tilde{\sigma}}$. Suppose we
have,%
\begin{equation}
\tilde{\partial}_{i}=E_{i}^{j}\partial _{j}\text{,}
\end{equation}%
where $E_{i}^{j}$ is a $2\times 2$ non singular matrix of real numbers
describing the basis transformation. A change of basis induces a change of
the components of a vector $\Theta $ in the tangent space $T_{p}\left( 
\mathcal{M}_{s}\right) $. Assume that $\tilde{\partial}_{i}=E_{i}^{j}%
\partial _{j}$ is a general basis transformation at $p$ and let $\Theta
=\Theta ^{i}\partial _{i}=\tilde{\Theta}^{i}\tilde{\partial}_{i}$ define the
components of a vector $\Theta $ with respect to the two bases $\mathcal{B}_{%
\text{old}}$ and $\mathcal{B}_{\text{new}}$ according to,%
\begin{equation}
\Theta =\Theta ^{i}\partial _{i}\overset{\text{def}}{=}\mu \partial
_{1}+\sigma \partial _{2}=\tilde{\Theta}^{i}\tilde{\partial}_{i}\overset{%
\text{def}}{=}\tilde{\mu}\tilde{\partial}_{1}+\tilde{\sigma}\tilde{\partial}%
_{2}\text{.}  \label{sop}
\end{equation}%
Noticing that $\tilde{\partial}_{i}=E_{i}^{j}\partial _{j}$ and using (\ref%
{sop}), we obtain 
\begin{equation}
\left( \tilde{\Theta}^{k}E_{k}^{i}-\Theta ^{i}\right) \partial _{i}=\left( 
\tilde{\Theta}^{i}E_{i}^{k}-\Theta ^{k}\right) \partial _{k}=0\text{,}
\end{equation}%
such that,%
\begin{equation}
\tilde{\Theta}^{i}=\left( E^{-1}\right) _{k}^{i}\Theta ^{k}\text{.}
\label{rr}
\end{equation}%
Equation (\ref{rr}) implies that the components of a vector change
controvariantly to the corresponding change of basis. As a final remark,
notice that%
\begin{equation}
\Theta =\Theta ^{i}\partial _{i}=\tilde{\Theta}^{i}\tilde{\partial}_{i}=%
\left[ \left( E^{-1}\right) _{j}^{i}\Theta ^{j}\right] \left[
E_{i}^{k}\partial _{k}\right] =\left( E^{-1}\right) _{j}^{i}E_{i}^{k}\Theta
^{j}\partial _{k}=\delta _{j}^{k}\Theta ^{j}\partial _{k}=\Theta
^{k}\partial _{k}\text{.}
\end{equation}%
These preliminary mathematical tools may be extended to the $2l$-dimensional
case in a fairly straightforward way.

\subsection{Diagonalization}

The information metric tensor $\hat{g}\left( \mu \text{, }\sigma \right) $
in (\ref{sm}) has a matrix representation in the basis $\mathcal{B}_{\text{%
old}}\overset{\text{def}}{=}\left\{ \partial _{1}\text{, }\partial
_{2}\right\} $ given by,%
\begin{equation}
\left[ \hat{g}\left( \mu \text{, }\sigma \right) \right] _{\mathcal{B}_{%
\text{old}}}\overset{\text{def}}{=}g_{ij}\left( \mu \text{, }\sigma \right) =%
\frac{1}{\sigma ^{2}}\left( 
\begin{array}{cc}
1 & r \\ 
r & 2%
\end{array}%
\right) \text{, with }i\text{, }j=1\text{, }2\text{.}
\end{equation}%
Since $g_{ij}\left( \mu \text{, }\sigma \right) $ is symmetric, it is
diagonalizable. The eigenvalues read\textbf{,}%
\begin{equation}
\alpha _{\pm }\left( r\right) \overset{\text{def}}{=}\frac{3\pm \sqrt{\Delta 
}}{2}\text{, }\Delta =1+4r^{2}\text{.}
\end{equation}%
As a side remark, notice that $\alpha _{-}\left( r\right) \overset{%
r\longrightarrow 0}{\longrightarrow }+1$ and, $\alpha _{+}\left( r\right) 
\overset{r\longrightarrow 0}{\longrightarrow }+2$. The eigenvectors $\Theta
_{+}\left( r\right) $ and $\Theta _{-}\left( r\right) $ corresponding to $%
\alpha _{+}\left( r\right) $ and $\alpha _{-}\left( r\right) $,
respectively, are,%
\begin{equation}
\Theta _{+}\left( r\right) =\left( 
\begin{array}{c}
1 \\ 
\frac{1+\sqrt{\Delta }}{2r}%
\end{array}%
\right) \text{, }\Theta _{-}\left( r\right) =\left( 
\begin{array}{c}
1 \\ 
\frac{1-\sqrt{\Delta }}{2r}%
\end{array}%
\right) \text{.}
\end{equation}%
Finally, the diagonalized information matrix $\left[ \hat{g}^{\prime }\left(
\mu \left( \tilde{\mu}\text{, }\tilde{\sigma}\right) \text{, }\sigma \left( 
\tilde{\mu}\text{, }\tilde{\sigma}\right) \right) \right] _{\mathcal{B}_{%
\text{new}}}$ in the new basis $\mathcal{B}_{\text{new}}$ satisfies the
following relation, 
\begin{equation}
\left[ \hat{g}\left( \mu \text{, }\sigma \right) \right] _{\mathcal{B}_{%
\text{old}}}=E\left( r\right) \left[ \hat{g}^{\prime }\left( \mu \left( 
\tilde{\mu}\text{, }\tilde{\sigma}\right) \text{, }\sigma \left( \tilde{\mu}%
\text{, }\tilde{\sigma}\right) \right) \right] _{\mathcal{B}_{\text{new}%
}}E^{-1}\left( r\right) \text{,}
\end{equation}%
where, explicitly we obtain%
\begin{equation}
\left[ \hat{g}^{\prime }\left( \mu \left( \tilde{\mu}\text{, }\tilde{\sigma}%
\right) \text{, }\sigma \left( \tilde{\mu}\text{, }\tilde{\sigma}\right)
\right) \right] _{\mathcal{B}_{\text{new}}}\overset{\text{def}}{=}\frac{1}{%
\sigma ^{2}\left( \tilde{\mu}\text{, }\tilde{\sigma}\right) }\left( 
\begin{array}{cc}
\frac{3-\sqrt{\Delta }}{2} & 0 \\ 
0 & \frac{3+\sqrt{\Delta }}{2}%
\end{array}%
\right) \text{.}
\end{equation}%
The columns of the matrix $E\left( r\right) $ encode the eigenvectors $%
\Theta _{+}\left( r\right) $ and $\Theta _{-}\left( r\right) $ of $\left[ 
\hat{g}\left( \mu \text{, }\sigma \right) \right] _{\mathcal{B}_{\text{old}%
}} $ and are given by,%
\begin{equation}
E\left( r\right) \overset{\text{def}}{=}\left( 
\begin{array}{cc}
1 & 1 \\ 
\frac{1-\sqrt{\Delta }}{2r} & \frac{1+\sqrt{\Delta }}{2r}%
\end{array}%
\right) \text{.}  \label{X}
\end{equation}%
For future convenience, we express the pair of macrovariables $\left( \mu 
\text{, }\sigma \right) $ in terms of the new statistical variables $\tilde{%
\mu}$ and $\tilde{\sigma}$,%
\begin{equation}
g_{\mu \nu }\left( \mu \text{, }\sigma \right) \overset{\text{diag}}{%
\longrightarrow }g_{\mu \nu }^{\prime }\left( \tilde{\mu}\text{, }\tilde{%
\sigma}\right) \text{.}
\end{equation}%
From differential geometry arguments, it follows that%
\begin{equation}
\left( 
\begin{array}{c}
\frac{\partial }{\partial \tilde{\mu}} \\ 
\frac{\partial }{\partial \tilde{\sigma}}%
\end{array}%
\right) =E\left( r\right) \left( 
\begin{array}{c}
\frac{\partial }{\partial \mu } \\ 
\frac{\partial }{\partial \sigma }%
\end{array}%
\right) \text{ and, }\left( 
\begin{array}{c}
\mu \\ 
\sigma%
\end{array}%
\right) =E\left( r\right) \left( 
\begin{array}{c}
\tilde{\mu} \\ 
\tilde{\sigma}%
\end{array}%
\right) \text{.}  \label{XX}
\end{equation}%
Substituting (\ref{X}) in (\ref{XX}), we finally obtain the formal relation
between the old $\left( \mu \text{, }\sigma \right) $ and new $\left( \tilde{%
\mu}\text{, }\tilde{\sigma}\right) $ set of macrovariables labelling the
macrostates $\Theta $ of our two-dimensional Gaussian statistical model in
presence of non-trivial off-diagonal terms,%
\begin{equation}
\mu \left( \tilde{\mu}\text{, }\tilde{\sigma}\right) \overset{\text{def}}{=}%
\tilde{\mu}+\tilde{\sigma}\text{ and, }\sigma \left( \tilde{\mu}\text{, }%
\tilde{\sigma}\right) \overset{\text{def}}{=}\frac{1-\sqrt{\Delta }}{2r}%
\tilde{\mu}+\frac{1+\sqrt{\Delta }}{2r}\text{ }\tilde{\sigma}\text{.}
\label{cambio}
\end{equation}%
Recall that $\Theta =\Theta ^{i}\partial _{i}=\tilde{\Theta}^{i}\tilde{%
\partial}_{i}$ with $\tilde{\partial}_{i}=E_{i}^{j}\partial _{j}$ and $%
\tilde{\theta}^{i}=\left( E^{-1}\right) _{k}^{i}\theta ^{k}$ where $E^{-1}$
is the inverse of (\ref{X}). Therefore, in our case we have%
\begin{eqnarray}
\Theta &=&\left[ \left( E^{-1}\right) _{1}^{1}E_{1}^{1}+\left( E^{-1}\right)
_{1}^{2}E_{2}^{1}\right] \Theta ^{1}\partial _{1}+\left[ \left(
E^{-1}\right) _{1}^{1}E_{1}^{2}+\left( E^{-1}\right) _{1}^{2}E_{2}^{2}\right]
\Theta ^{1}\partial _{2}  \notag \\
&&  \label{XXX} \\
&&+\left[ \left( E^{-1}\right) _{2}^{1}E_{1}^{1}+\left( E^{-1}\right)
_{2}^{2}E_{2}^{1}\right] \Theta ^{2}\partial _{1}+\left[ \left(
E^{-1}\right) _{2}^{1}E_{1}^{2}+\left( E^{-1}\right) _{2}^{2}E_{2}^{2}\right]
\Theta ^{2}\partial _{2}\text{.}  \notag
\end{eqnarray}%
Moreover, notice that,%
\begin{equation}
\left( 
\begin{array}{cc}
E_{1}^{1} & E_{2}^{1} \\ 
E_{1}^{2} & E_{2}^{2}%
\end{array}%
\right) =\left( 
\begin{array}{cc}
1 & 1 \\ 
\frac{1-\sqrt{\Delta }}{2r} & \frac{1+\sqrt{\Delta }}{2r}%
\end{array}%
\right) \text{ and, }\left( 
\begin{array}{cc}
\left( E^{-1}\right) _{1}^{1} & \left( E^{-1}\right) _{2}^{1} \\ 
\left( E^{-1}\right) _{1}^{2} & \left( E^{-1}\right) _{2}^{2}%
\end{array}%
\right) =\frac{r}{\sqrt{\Delta }}\left( 
\begin{array}{cc}
\frac{1+\sqrt{\Delta }}{2r} & -1 \\ 
-\frac{1-\sqrt{\Delta }}{2r} & 1%
\end{array}%
\right) \text{.}  \label{X1}
\end{equation}%
Therefore, substituting (\ref{X1}) in (\ref{XXX}), we obtain,%
\begin{equation}
\Theta =\Theta ^{i}\partial _{i}=\tilde{\Theta}^{i}\tilde{\partial}_{i}\text{
with }\tilde{\partial}_{i}=E_{i}^{j}\partial _{j},\tilde{\Theta}^{i}=\left(
E^{-1}\right) _{k}^{i}\Theta ^{k}\text{.}
\end{equation}%
This calculation completes the verification of the correctness of our
information geometric diagonalization procedure.

\section{Information Geometric Complexity}

In this Section, we compute the asymptotic temporal behavior of the
dynamical complexity of the geodesic trajectories for the correlated
two-dimensional Gaussian statistical model.

\subsection{Integration of the Geodesic Equations}

After having introduced the diagonalization procedure, the new line element $%
ds^{\prime 2}\left( \tilde{\mu}\text{, }\tilde{\sigma}\right) $ to be
considered becomes,%
\begin{equation}
ds^{\prime 2}\left( \tilde{\mu}\text{, }\tilde{\sigma}\right) =\frac{\alpha
_{-}\left( r\right) }{\left[ a_{0}\left( r\right) \tilde{\mu}+a_{1}\left(
r\right) \tilde{\sigma}\right] ^{2}}d\tilde{\mu}^{2}+\frac{\alpha _{+}\left(
r\right) }{\left[ a_{0}\left( r\right) \tilde{\mu}+a_{1}\left( r\right) 
\tilde{\sigma}\right] ^{2}}d\tilde{\sigma}^{2}\text{,}
\end{equation}%
where,%
\begin{equation}
\alpha _{\pm }\left( r\right) \overset{\text{def}}{=}\frac{3\pm \sqrt{\Delta 
}}{2}\text{, }a_{0}\left( r\right) \overset{\text{def}}{=}\frac{1-\sqrt{%
\Delta }}{2r}\text{, }a_{1}\left( r\right) \overset{\text{def}}{=}\frac{1+%
\sqrt{\Delta }}{2r}\text{ and, }\Delta \left( r\right) \overset{\text{def}}{=%
}1+4r^{2}\text{.}
\end{equation}%
Notice that $ds^{\prime 2}\left( \tilde{\mu}\text{, }\tilde{\sigma}\right) $
can be rewritten as,%
\begin{equation}
ds^{\prime 2}\left( \tilde{\mu}\text{, }\tilde{\sigma}\right) =\frac{\alpha
_{-}\left( r\right) }{\left[ a_{1}\left( r\right) \right] ^{2}}\frac{1}{%
\tilde{\sigma}^{2}}\frac{1}{\left( 1+\frac{a_{0}\left( r\right) }{%
a_{1}\left( r\right) }\frac{\tilde{\mu}}{\tilde{\sigma}}\right) ^{2}}d\tilde{%
\mu}^{2}+\frac{\alpha _{+}\left( r\right) }{\left[ a_{1}\left( r\right) %
\right] ^{2}}\frac{1}{\tilde{\sigma}^{2}}\frac{1}{\left( 1+\frac{a_{0}\left(
r\right) }{a_{1}\left( r\right) }\frac{\tilde{\mu}}{\tilde{\sigma}}\right)
^{2}}d\tilde{\sigma}^{2}\text{.}
\end{equation}%
As a working hypothesis, we assume that $\frac{a_{0}\left( r\right) }{%
a_{1}\left( r\right) }\frac{\tilde{\mu}\left( \tau \right) }{\tilde{\sigma}%
\left( \tau \right) }\ll 1$ for $\tau \gg 1$. Stated otherwise, we assume
that%
\begin{equation}
\lim_{\tau \rightarrow \infty }\left[ \frac{\tilde{\mu}\left( \tau \right) }{%
\tilde{\sigma}\left( \tau \right) }\right] \ll \min_{r\in \left( 0\text{,}%
1\right) }\left\vert \frac{a_{1}\left( r\right) }{a_{0}\left( r\right) }%
\right\vert =\min_{r\in \left( 0\text{,}1\right) }\left\vert \frac{1+\sqrt{%
1+4r^{2}}}{1-\sqrt{1+4r^{2}}}\right\vert \simeq 2.6\text{.}  \label{1}
\end{equation}%
Then, in the\textit{\ long time limit}, the notion of distinguishability
between probability distributions on the diagonalized statistical manifold
is quantified by the following line element,%
\begin{equation}
ds^{\prime 2}\left( \tilde{\mu}\text{, }\tilde{\sigma}\right) =\frac{\alpha
_{-}\left( r\right) }{\left[ a_{1}\left( r\right) \right] ^{2}}\frac{1}{%
\tilde{\sigma}^{2}}d\tilde{\mu}^{2}+\frac{\alpha _{+}\left( r\right) }{\left[
a_{1}\left( r\right) \right] ^{2}}\frac{1}{\tilde{\sigma}^{2}}d\tilde{\sigma}%
^{2}\text{.}  \label{qq}
\end{equation}%
After computing the Christoffel connection coefficients arising from (\ref%
{qq}), the set of coupled nonlinear ordinary differential equations
satisfied by the geodesic trajectories becomes,%
\begin{equation}
\frac{d^{2}\tilde{\mu}}{d\tau ^{2}}-\frac{2}{\tilde{\sigma}}\frac{d\tilde{\mu%
}}{d\tau }\frac{d\tilde{\sigma}}{d\tau }=0\text{, }\frac{d^{2}\tilde{\sigma}%
}{d\tau ^{2}}+\frac{\alpha _{-}\left( r\right) }{\alpha _{+}\left( r\right) }%
\frac{1}{\tilde{\sigma}}\left( \frac{d\tilde{\mu}}{d\tau }\right) ^{2}-\frac{%
1}{\tilde{\sigma}}\left( \frac{d\tilde{\sigma}}{d\tau }\right) ^{2}=0\text{.}
\label{ss}
\end{equation}%
Notice that in the limit of $r\rightarrow 0$, $\frac{\alpha _{-}\left(
r\right) }{\alpha _{+}\left( r\right) }\rightarrow \frac{1}{2}$ and the
system of equations (\ref{ss}) describing the asymptotic behavior of
maximally probable trajectories on the diagonalized manifold reduces to the
standard two-dimensional Gaussian system of nonlinear coupled ordinary
differential equations studied in \cite{cafaroPD}. In order to further
simply the integration of (\ref{ss}), consider the following (invertible)
change of variables,%
\begin{equation}
\left( \tilde{\mu}\text{, }\tilde{\sigma}\right) \longrightarrow \left( \mu
^{\prime }\left( \tilde{\mu}\text{, }\tilde{\sigma}\right) =\sqrt{\frac{%
2\alpha _{-}\left( r\right) }{\alpha _{+}\left( r\right) }}\tilde{\mu}\text{%
, }\sigma ^{\prime }\left( \tilde{\mu}\text{, }\tilde{\sigma}\right) =\tilde{%
\sigma}\right) \text{. }  \label{33}
\end{equation}%
Substituting (\ref{33}) into (\ref{ss}), the coupled system of nonlinear
differential equations to be integrated becomes,%
\begin{equation}
\frac{d^{2}\mu ^{\prime }}{d\tau ^{2}}-\frac{2}{\sigma ^{\prime }}\frac{d\mu
^{\prime }}{d\tau }\frac{d\sigma ^{\prime }}{d\tau }=0\text{, }\frac{%
d^{2}\sigma ^{\prime }}{d\tau ^{2}}+\frac{1}{2\sigma ^{\prime }}\left( \frac{%
d\mu ^{\prime }}{d\tau }\right) ^{2}-\frac{1}{\sigma ^{\prime }}\left( \frac{%
d\sigma ^{\prime }}{d\tau }\right) ^{2}=0\text{.}  \label{ddd}
\end{equation}%
Integrating (\ref{ddd}) leads to the following geodesic trajectories,%
\begin{equation}
\mu ^{\prime }\left( \tau \right) =\frac{\Xi ^{2}}{2\lambda }\frac{1}{\exp
\left( -2\lambda \tau \right) +\frac{\Xi ^{2}}{8\lambda ^{2}}}-4\lambda 
\text{, }\sigma ^{\prime }\left( \tau \right) =\frac{\Xi \exp \left(
-\lambda \tau \right) }{\exp \left( -2\lambda \tau \right) +\frac{\Xi ^{2}}{%
8\lambda ^{2}}}\text{,}
\end{equation}%
where $\Xi $ and $\lambda $ are \textit{real} and \textit{positive}
constants of integration \cite{cafaroPD}. Using (\ref{cambio}) and (\ref{33}%
), we have%
\begin{equation}
\mu \left( \mu ^{\prime }\text{, }\sigma ^{\prime }\right) \overset{\text{def%
}}{=}\sqrt{\frac{\alpha _{+}\left( r\right) }{2\alpha _{-}\left( r\right) }}%
\mu ^{\prime }+\sigma ^{\prime }\text{ and, }\sigma \left( \mu ^{\prime }%
\text{, }\sigma ^{\prime }\right) \overset{\text{def}}{=}\frac{1-\sqrt{%
\Delta }}{2r}\sqrt{\frac{\alpha _{+}\left( r\right) }{2\alpha _{-}\left(
r\right) }}\mu ^{\prime }+\frac{1+\sqrt{\Delta }}{2r}\sigma ^{\prime }\text{.%
}
\end{equation}%
Notice that our working hypothesis (\ref{1}) is satisfied since we have,%
\begin{equation}
\lim_{\tau \rightarrow \infty }\frac{\tilde{\mu}\left( \tau \right) }{\tilde{%
\sigma}\left( \tau \right) }=\sqrt{\frac{\alpha _{+}\left( r\right) }{%
2\alpha _{-}\left( r\right) }}\lim_{\tau \rightarrow \infty }\frac{\mu
^{\prime }\left( \tau \right) }{\sigma ^{\prime }\left( \tau \right) }%
\propto \exp \left( -\lambda \tau \right) \overset{\tau \rightarrow \infty }{%
\longrightarrow }0\text{.}
\end{equation}%
Finally, in terms of the original macrovariables $\left( \mu \text{, }\sigma
\right) $, the geodesic trajectories finally become, 
\begin{equation}
\begin{array}{c}
\mu \left( \tau \text{; }r\right) =\sqrt{\frac{\alpha _{+}\left( r\right) }{%
2\alpha _{-}\left( r\right) }}\left[ \frac{\Xi ^{2}}{2\lambda }\frac{1}{\exp
\left( -2\lambda \tau \right) +\frac{\Xi ^{2}}{8\lambda ^{2}}}-4\lambda %
\right] +\frac{\Xi \exp \left( -\lambda \tau \right) }{\exp \left( -2\lambda
\tau \right) +\frac{\Xi ^{2}}{8\lambda ^{2}}}\text{,} \\ 
\\ 
\sigma \left( \tau \text{; }r\right) =\frac{1-\sqrt{\Delta \left( r\right) }%
}{2r}\sqrt{\frac{\alpha _{+}\left( r\right) }{2\alpha _{-}\left( r\right) }}%
\left[ \frac{\Xi ^{2}}{2\lambda }\frac{1}{\exp \left( -2\lambda \tau \right)
+\frac{\Xi ^{2}}{8\lambda ^{2}}}-4\lambda \right] +\frac{1+\sqrt{\Delta
\left( r\right) }}{2r}\frac{\Xi \exp \left( -\lambda \tau \right) }{\exp
\left( -2\lambda \tau \right) +\frac{\Xi ^{2}}{8\lambda ^{2}}}\text{.}%
\end{array}
\label{GGE}
\end{equation}%
In our probabilistic macroscopic approach to dynamics, the geodesic
trajectories in (\ref{GGE}) represent the maximum probability paths for the
Gaussian statistical model in the presence of macro-correlations.

\subsection{Computation of the Information Geometric Complexity}

In our information geometric approach a relevant quantity that can be useful
to study the degree of complexity characterizing information-constrained
dynamical models is the information geometrodynamical entropy $\mathcal{S}_{%
\mathcal{M}_{s}}\left( \tau \right) $ (IGE) defined as \cite{cafaroPD},%
\begin{equation}
\mathcal{S}_{\mathcal{M}_{s}}\left( \tau \right) \overset{\text{def}}{=}%
\underset{\tau \rightarrow \infty }{\lim }\log \mathcal{V}_{\mathcal{M}%
_{s}}\left( \tau \right) =\underset{\tau \rightarrow \infty }{\lim }\log %
\left[ \frac{1}{\tau }\dint\limits_{0}^{\tau }d\tau ^{\prime }\left( 
\underset{\Theta _{i}\left( 0\right) }{\overset{\Theta _{f}\left( \tau
^{\prime }\right) }{\int }}\sqrt{g}d^{2l}\Theta \right) \right] \text{, }
\label{info}
\end{equation}%
and $g=\left\vert \det \left( g_{\mu \nu }\right) \right\vert $. The IGE is
expressed in terms of the information geometric complexity $\mathcal{V}_{%
\mathcal{M}_{s}}\left( \tau \right) $. It is the asymptotic limit of the
natural logarithm of the statistical weight defined on the $2l$-dimensional $%
\mathcal{M}_{s}$ and represents a measure of temporal complexity of chaotic
dynamical systems whose dynamics is underlined by a curved statistical
manifold. In our case $l=1$ and $\mathcal{V}_{\mathcal{M}_{s}}\left( \tau
\right) $ becomes,%
\begin{equation}
\mathcal{V}_{\mathcal{M}_{s}}\left( \tau \right) =\frac{1}{\tau }%
\dint\limits_{0}^{\tau }d\tau ^{\prime }\left( \underset{\Theta _{i}\left(
0\right) }{\overset{\Theta _{f}\left( \tau ^{\prime }\right) }{\int }}\sqrt{g%
}d^{2}\Theta \right) \text{,}  \label{q2}
\end{equation}%
where $g\left( r\right) =\frac{2-r^{2}}{\sigma ^{2}}$ and $\Theta \left(
\tau \right) =\left( \mu \left( \tau \text{; }r\right) \text{, }\sigma
\left( \tau \text{; }r\right) \text{ }\right) $ is given in (\ref{GGE}).
Substituting $\sqrt{g}$ and $\Theta \left( \tau \right) $ into (\ref{q2}),
we obtain,%
\begin{equation}
\mathcal{V}_{\mathcal{M}_{s}}\left( \tau \right) =\frac{\sqrt{2-r^{2}}}{\tau 
}\dint\limits^{\tau }\left[ \frac{\Xi \exp \left( -\lambda \tau ^{\prime
}\right) -4\lambda \sqrt{\frac{\alpha _{+}\left( r\right) }{2\alpha
_{-}\left( r\right) }}\exp \left( -2\lambda \tau ^{\prime }\right) }{\frac{1+%
\sqrt{\Delta \left( r\right) }}{2r}\Xi \exp \left( -\lambda \tau ^{\prime
}\right) -4\lambda \sqrt{\frac{\alpha _{+}\left( r\right) }{2\alpha
_{-}\left( r\right) }}\frac{1-\sqrt{\Delta \left( r\right) }}{2r}\exp \left(
-2\lambda \tau ^{\prime }\right) }\right] d\tau ^{\prime }\text{.}
\end{equation}%
For the sake of simplicity, let us introduce the following substitutions,%
\begin{equation}
A\overset{\text{def}}{=}\Xi \text{, }B\overset{\text{def}}{=}-4\lambda \sqrt{%
\frac{\alpha _{+}\left( r\right) }{2\alpha _{-}\left( r\right) }}\text{, }C%
\overset{\text{def}}{=}\frac{1+\sqrt{\Delta \left( r\right) }}{2r}\Xi \text{%
, }D\overset{\text{def}}{=}-4\lambda \sqrt{\frac{\alpha _{+}\left( r\right) 
}{2\alpha _{-}\left( r\right) }}\frac{1-\sqrt{\Delta \left( r\right) }}{2r}%
\text{.}  \label{casa}
\end{equation}%
Then, the integral defining $\mathcal{V}_{\mathcal{M}_{s}}\left( \tau
\right) $ becomes,%
\begin{equation}
\mathcal{V}_{\mathcal{M}_{s}}\left( \tau \right) =\frac{\sqrt{2-r^{2}}}{\tau 
}\dint\limits^{\tau }\left[ \frac{Ae^{-\lambda \tau ^{\prime
}}+Be^{-2\lambda \tau ^{\prime }}}{Ce^{-\lambda \tau ^{\prime
}}+De^{-2\lambda \tau ^{\prime }}}d\tau ^{\prime }\right] d\tau ^{\prime }%
\text{.}  \label{29}
\end{equation}%
Upon integration, we get%
\begin{equation}
\int^{\tau }\frac{Ae^{-\lambda \tau ^{\prime }}+Be^{-2\lambda \tau ^{\prime
}}}{Ce^{-\lambda \tau ^{\prime }}+De^{-2\lambda \tau ^{\prime }}}d\tau
^{\prime }=\allowbreak \frac{1}{\lambda }\left( \frac{A}{C}-\frac{B}{D}%
\right) \ln \left[ \frac{D+Ce^{\lambda \tau }}{De^{\lambda \tau }}\right] +%
\frac{A}{C}\tau \overset{\tau \rightarrow \infty }{\approx }\allowbreak 
\frac{1}{\lambda }\left( \frac{A}{C}-\frac{B}{D}\right) \ln \frac{C}{D}+%
\frac{A}{C}\tau \text{.}  \label{17}
\end{equation}%
Upon substituting (\ref{17}) into (\ref{29}), we obtain%
\begin{equation}
\mathcal{V}_{\mathcal{M}_{s}}\left( \tau \right) =\sqrt{2-r^{2}}\left[ \frac{%
A}{C}+\frac{1}{\lambda }\left( \frac{A}{C}-\frac{B}{D}\right) \ln \frac{C}{D}%
\frac{1}{\tau }\right] \text{.}
\end{equation}%
By re-introducing the original parameters in (\ref{casa}), we get%
\begin{equation}
\mathcal{V}_{\mathcal{M}_{s}}\left( \tau \right) =\frac{2r\sqrt{2-r^{2}}}{1+%
\sqrt{\Delta \left( r\right) }}+\left[ \frac{2r\sqrt{2-r^{2}}}{\left( 1+%
\sqrt{\Delta \left( r\right) }\right) \lambda }\ln \Sigma \left( r\text{, }%
\lambda \text{, }\alpha _{\pm }\right) -\frac{2r\sqrt{2-r^{2}}}{\left( 1-%
\sqrt{\Delta \left( r\right) }\right) \lambda }\ln \Sigma \left( r\text{, }%
\lambda \text{, }\alpha _{\pm }\right) \right] \frac{1}{\tau }\text{,}
\label{yes}
\end{equation}%
where the strictly positive function $\Sigma \left( r\text{, }\lambda \text{%
, }\alpha _{\pm }\right) $ is given by,%
\begin{equation}
\Sigma \left( r\text{, }\lambda \text{, }\alpha _{\pm }\right) \overset{%
\text{def}}{=}\frac{\frac{1+\sqrt{\Delta \left( r\right) }}{2r}\Xi }{%
-4\lambda \sqrt{\frac{\alpha _{+}\left( r\right) }{2\alpha _{-}\left(
r\right) }}\frac{1-\sqrt{\Delta \left( r\right) }}{2r}}>0\text{, }\forall
r\in \left( 0\text{, }1\right) \text{.}
\end{equation}%
Finally, inserting (\ref{yes}) into (\ref{info}), the IGE $\mathcal{S}_{%
\mathcal{M}_{s}}\left( \tau \right) $ becomes,%
\begin{equation}
\mathcal{S}_{\mathcal{M}_{s}}\left( \tau \text{; }\lambda \text{, }r\right)
=\log \left\{ \frac{2r\sqrt{2-r^{2}}}{1+\sqrt{\Delta \left( r\right) }}+%
\left[ \frac{2r\sqrt{2-r^{2}}\ln \Sigma \left( r\text{, }\lambda \text{, }%
\alpha _{\pm }\right) }{\left( 1+\sqrt{\Delta \left( r\right) }\right)
\lambda }-\frac{2r\sqrt{2-r^{2}}\ln \Sigma \left( r\text{, }\lambda \text{, }%
\alpha _{\pm }\right) }{\left( 1-\sqrt{\Delta \left( r\right) }\right)
\lambda }\right] \frac{1}{\tau }\right\} \text{.}  \label{ige}
\end{equation}%
With a suitable change of notation, equation (\ref{ige})\textbf{\ }can be
rewritten more elegantly as,%
\begin{equation}
\mathcal{S}_{\mathcal{M}_{s}}\left( \tau \text{; }\lambda \text{, }r\right) 
\overset{\tau \rightarrow \infty }{\sim }\log \left[ \Lambda _{1}\left(
r\right) +\frac{\Lambda _{2}\left( r\text{, }\lambda \right) }{\tau }\right] 
\text{,}
\end{equation}%
where,%
\begin{equation}
\Lambda _{1}\left( r\right) \overset{\text{def}}{=}\frac{2r\sqrt{2-r^{2}}}{1+%
\sqrt{1+4r^{2}}}\text{, }\Lambda _{2}\left( r\text{, }\lambda \right) 
\overset{\text{def}}{=}\frac{\sqrt{\left( 1+4r^{2}\right) \left(
2-r^{2}\right) }}{r}\frac{\ln \Sigma \left( r\text{, }\lambda \text{, }%
\alpha _{\pm }\right) }{\lambda }\text{, }\alpha _{\pm }\left( r\right) 
\overset{\text{def}}{=}\frac{3\pm \sqrt{1+4r^{2}}}{2}\text{. }
\end{equation}%
As stated above\textbf{,} $\Sigma \left( r\text{, }\lambda \text{, }\alpha
_{\pm }\right) $ is a strictly positive function of its arguments. It
appears that the introduction of additional information constraints
(correlations; non-trivial off-diagonal terms) between the macrovariables of
the Gaussian statistical model leads to the emergence of an asymptotic
information geometric compression of the statistical macrostates explored on
the configuration manifold of the Gaussian statistical model $\mathcal{M}%
_{s} $ in its evolution between the initial and final macrostates. This
result, although valid for a special (not general) case, leads to
interesting conclusions. The presence of correlations between macroscopic
information about the microscopic degrees of freedom of a complex system
allows for an information geometric probabilistic description whose
complexity, measured in terms of\textbf{\ }$\mathcal{S}_{\mathcal{M}_{s}}$%
\textbf{\ (}or\textbf{\ }$\mathcal{V}_{\mathcal{M}_{s}}$\textbf{)}, decays
according to a power law. Asymptotically, the complexity reaches a
saturation value characterized solely by the strength of such correlations.
The relevance of such results becomes more transparent when compared to what
occurs in absence of correlations \cite{cafaroPD}. We will mention this
comparison in our final remarks.

\section{The $2l$-Dimensional Macro-correlated Gaussian Statistical Model}

In this Section, we outline the main steps involved in the computation of
the asymptotic temporal behavior of the dynamical complexity of geodesic
trajectories for the correlated $2l$-dimensional Gaussian statistical model.
Recall that the line element $ds^{2}=g_{ij}\left( \Theta \right) d\Theta
^{i}d\Theta ^{j}$ on the $2l$-dimensional Gaussian statistical manifold $%
\mathcal{M}_{s}$ in the presence of non trivial off-diagonal terms is given
by,%
\begin{equation}
ds_{\mathcal{M}_{s}}^{2}\overset{\text{def}}{=}\sum_{j=1}^{l}\left( \frac{1}{%
\sigma _{j}^{2}}d\mu _{j}^{2}+\frac{2r_{j}}{\sigma _{j}^{2}}d\mu _{j}d\sigma
_{j}+\frac{2}{\sigma _{j}^{2}}d\sigma _{j}^{2}\right) \text{, with }\mu 
\text{, }\nu =1\text{,..., }2l\text{.}  \label{SM}
\end{equation}%
We consider positive coefficients $r_{j}\in \left( 0\text{, }1\right) $, $%
\forall j=1$,..., $l$. The Fisher-Rao metric tensor $g_{ij}\left( \Theta
\right) \overset{\text{def}}{=}g_{ij}\left( \mu _{1}\text{,...}\mu _{l}\text{%
; }\sigma _{1}\text{,..., }\sigma _{l}\right) $ leading to the line element
in (\ref{SM}) is given by,%
\begin{equation}
\left[ g_{ij}\left( \Theta \right) \right] _{2l\times 2l}=\left( 
\begin{array}{cccc}
M_{2\times 2}^{\left( 1\right) } & 0 & 0 & 0 \\ 
0 & \cdot & 0 & 0 \\ 
0 & 0 & \cdot & 0 \\ 
0 & 0 & 0 & M_{2\times 2}^{\left( l\right) }%
\end{array}%
\right) \text{, with }i\text{, }j=1\text{,..., }2l\text{.}
\end{equation}%
where $M_{2\times 2}^{\left( k\right) }$ is the two-dimensional matrix
defined as,%
\begin{equation}
\left[ M_{2\times 2}^{\left( k\right) }\right] \overset{\text{def}}{=}\frac{1%
}{\sigma _{k}^{2}}\left( 
\begin{array}{cc}
1 & r_{k} \\ 
r_{k} & 2%
\end{array}%
\right) \text{ with }k=1\text{,..., }l\text{.}
\end{equation}%
The inverse matrix $\left[ M_{2\times 2}^{\left( k\right) }\right] ^{-1}$,
useful for computing the Christoffel connection coefficients and other
quantities characterizing the information geometry of $\mathcal{M}_{s}$ is
given by,%
\begin{equation}
\left[ M_{2\times 2}^{\left( k\right) }\right] ^{-1}\overset{\text{def}}{=}%
\frac{\sigma _{k}^{2}}{2-r_{k}^{2}}\left( 
\begin{array}{cc}
2 & -r_{k} \\ 
-r_{k} & 1%
\end{array}%
\right) \text{ with }k=1\text{,..., }l\text{.}
\end{equation}%
Following Section III, it can be shown that the Ricci scalar curvature of
such $2l$-dimensional manifold is given by,%
\begin{equation}
\mathcal{R}_{\mathcal{M}_{s}}\left( r_{1}\text{,..., }r_{l}\right)
=-2\sum_{k=1}^{l}\left( 2-r_{k}^{2}\right) ^{-1}\text{.}
\end{equation}%
Notice that in the limit of vanishing correlation strengths $\left\{
r_{k}\right\} $, $\mathcal{R}_{\mathcal{M}_{s}}=-l$ as shown in \cite%
{cafaroIJTP}. The computation of geodesic equations on the $2l$-dimensional
Gaussian statistical manifold $\mathcal{M}_{s}$ leads to the following
coupled systems of nonlinear second order ordinary differential equations,%
\begin{eqnarray}
0 &=&\frac{d^{2}\mu _{k}}{d\tau ^{2}}-\frac{r_{k}}{2-r_{k}^{2}}\frac{1}{%
\sigma _{k}}\left( \frac{d\mu _{k}}{d\tau }\right) ^{2}-\frac{4}{2-r_{k}^{2}}%
\frac{1}{\sigma _{k}}\frac{d\mu _{k}}{d\tau }\frac{d\sigma _{k}}{d\tau }-%
\frac{2r_{k}}{2-r_{k}^{2}}\frac{1}{\sigma _{k}}\left( \frac{d\sigma _{k}}{%
d\tau }\right) ^{2}\text{,}  \notag \\
&&  \notag \\
0 &=&\frac{d^{2}\sigma _{k}}{d\tau ^{2}}+\frac{1}{2-r_{k}^{2}}\frac{1}{%
\sigma _{k}}\left( \frac{d\mu _{k}}{d\tau }\right) ^{2}+\frac{2r_{k}}{%
2-r_{k}^{2}}\frac{1}{\sigma _{k}}\frac{d\mu _{k}}{d\tau }\frac{d\sigma _{k}}{%
d\tau }+\frac{2r_{k}^{2}-2}{2-r_{k}^{2}}\frac{1}{\sigma _{k}}\left( \frac{%
d\sigma _{k}}{d\tau }\right) ^{2}\text{.}
\end{eqnarray}%
with $k=1$,..., $l$. When $r_{k}\rightarrow 0$, $\forall k$ we get the
ordinary Gaussian system of nonlinear and coupled ordinary differential
equations. Integration of such coupled system of nonlinear second order
ordinary differential equations in is highly non trivial as pointed out in
Section III. However, this problem can be tackled extending the information
geometric diagonalization procedure introduced in Section IV to the\textit{\ 
}$2l$-dimensional case. The information metric tensor $\hat{g}\left( \mu _{1}%
\text{,...}\mu _{l}\text{; }\sigma _{1}\text{,..., }\sigma _{l}\right) 
\overset{\text{def}}{=}\hat{g}\left( \Theta \right) $ in (\ref{SM}) is
symmetric and therefore diagonalizable. The eigenvalues of such matrix are,%
\begin{equation}
\alpha _{\pm }\left( r_{k}\right) \overset{\text{def}}{=}\frac{3\pm \sqrt{%
\Delta \left( r_{k}\right) }}{2}\text{, }\Delta \left( r_{k}\right)
=1+4r_{k}^{2}\text{, with }k=1\text{,..., }l\text{.}
\end{equation}%
The eigenvectors $\Theta _{+}^{\left( k\right) }\overset{\text{def}}{=}%
\Theta _{+}\left( r_{k}\right) $ and $\Theta _{-}^{\left( k\right) }\overset{%
\text{def}}{=}\Theta _{-}\left( r\right) $ corresponding to $\alpha
_{+}\left( r_{k}\right) $ and $\alpha _{-}\left( r_{k}\right) $,
respectively, are,%
\begin{equation}
\Theta _{+}\left( r_{k}\right) =\left( 
\begin{array}{c}
1 \\ 
\frac{1+\sqrt{\Delta \left( r_{k}\right) }}{2r_{k}}%
\end{array}%
\right) \text{and, }\Theta _{-}\left( r_{k}\right) =\left( 
\begin{array}{c}
1 \\ 
\frac{1-\sqrt{\Delta \left( r_{k}\right) }}{2r_{k}}%
\end{array}%
\right) \text{ with }k=1\text{,..., }l\text{. }
\end{equation}%
Following the notation introduced in Section IV, the diagonalized
information matrix $\left[ \hat{g}^{\prime }\left( \Theta \left( \tilde{%
\Theta}\right) \right) \right] _{\mathcal{B}_{\text{new}}}$ in the new basis 
$\mathcal{B}_{\text{new}}$ satisfies the following relation, 
\begin{equation}
\left[ \hat{g}\left( \Theta \right) \right] _{\mathcal{B}_{\text{old}%
}}=E_{2l\times 2l}\left( r_{1}\text{,..., }r_{l}\right) \left[ \hat{g}%
^{\prime }\left( \Theta \left( \tilde{\Theta}\right) \right) \right] _{%
\mathcal{B}_{\text{new}}}E_{2l\times 2l}^{-1}\left( r_{1}\text{,..., }%
r_{l}\right) \text{,}
\end{equation}%
where, in an explicit way, we obtain%
\begin{equation}
\left[ \hat{g}^{\prime }\left( \Theta \left( \tilde{\Theta}\right) \right) %
\right] _{\mathcal{B}_{\text{new}}}=\left( 
\begin{array}{cccc}
D_{2\times 2}^{\left( 1\right) } & 0 & 0 & 0 \\ 
0 & \cdot & 0 & 0 \\ 
0 & 0 & \cdot & 0 \\ 
0 & 0 & 0 & D_{2\times 2}^{\left( l\right) }%
\end{array}%
\right) \text{,}
\end{equation}%
with the two-dimensional diagonal matrices $D_{2\times 2}^{\left( k\right) }$
defined as,%
\begin{equation}
\left[ D_{2\times 2}^{\left( k\right) }\right] \overset{\text{def}}{=}\frac{1%
}{\sigma _{k}^{2}\left( \tilde{\mu}_{k}\text{, }\tilde{\sigma}_{k}\right) }%
\left( 
\begin{array}{cc}
\frac{3-\sqrt{\Delta \left( r_{k}\right) }}{2} & 0 \\ 
0 & \frac{3+\sqrt{\Delta \left( r_{k}\right) }}{2}%
\end{array}%
\right) \text{ with }k=1\text{,..., }l\text{.}
\end{equation}%
The columns of the matrix $E_{2l\times 2l}\left( r_{1}\text{,..., }%
r_{l}\right) $ encode the eigenvectors $\Theta _{+}^{\left( k\right) }$ and $%
\Theta _{-}^{\left( k\right) }$ of $\left[ \hat{g}\left( \Theta \right) %
\right] _{\mathcal{B}_{\text{old}}}$\textbf{, }and are given by,%
\begin{equation}
\left[ E\left( r_{1}\text{,..., }r_{l}\right) \right] _{2l\times 2l}=\left( 
\begin{array}{cccc}
E_{2\times 2}^{\left( 1\right) } & 0 & 0 & 0 \\ 
0 & \cdot & 0 & 0 \\ 
0 & 0 & \cdot & 0 \\ 
0 & 0 & 0 & E_{2\times 2}^{\left( l\right) }%
\end{array}%
\right) \text{,}
\end{equation}%
where the two-dimensional matrices $E_{2\times 2}^{\left( k\right) }$ are,%
\begin{equation}
\left[ E_{2\times 2}^{\left( k\right) }\right] \overset{\text{def}}{=}\left( 
\begin{array}{cc}
1 & 1 \\ 
\frac{1-\sqrt{\Delta \left( r_{k}\right) }}{2r_{k}} & \frac{1+\sqrt{\Delta
\left( r_{k}\right) }}{2r_{k}}%
\end{array}%
\right) \text{.}  \label{ZZ}
\end{equation}%
The utility of $E_{2l\times 2l}\left( r_{1}\text{,..., }r_{l}\right) $ (and
its inverse) rests in its ability to express the set of macrovariables $%
\left( \mu _{1}\text{,...}\mu _{l}\text{; }\sigma _{1}\text{,..., }\sigma
_{l}\right) $ in terms of the new statistical variables $\left( \tilde{\mu}%
_{1}\text{,..., }\tilde{\mu}_{l}\text{; }\tilde{\sigma}_{1}\text{,..., }%
\tilde{\sigma}_{l}\right) $,%
\begin{equation}
g_{ij}\left( \mu _{1}\text{,...}\mu _{l}\text{; }\sigma _{1}\text{,..., }%
\sigma _{l}\right) \overset{\text{diag}}{\longrightarrow }g_{ij}^{\prime
}\left( \tilde{\mu}_{1}\text{,..., }\tilde{\mu}_{l}\text{; }\tilde{\sigma}%
_{1}\text{,..., }\tilde{\sigma}_{l}\right) \text{.}
\end{equation}%
From differential geometry arguments, it follows that%
\begin{equation}
\left( 
\begin{array}{c}
\partial _{\tilde{\mu}_{1}} \\ 
\cdot \\ 
\cdot \\ 
\partial _{\tilde{\sigma}_{l}}%
\end{array}%
\right) =E_{2l\times 2l}\left( r_{1}\text{,..., }r_{l}\right) \left( 
\begin{array}{c}
\partial _{\mu _{1}} \\ 
\cdot \\ 
\cdot \\ 
\partial _{\sigma _{l}}%
\end{array}%
\right) \text{ and, }\left( 
\begin{array}{c}
\mu _{1} \\ 
\cdot \\ 
\cdot \\ 
\sigma _{l}%
\end{array}%
\right) =E_{2l\times 2l}\left( r_{1}\text{,..., }r_{k}\right) \left( 
\begin{array}{c}
\tilde{\mu}_{1} \\ 
\cdot \\ 
\cdot \\ 
\tilde{\sigma}_{l}%
\end{array}%
\right) \text{.}  \label{ZZZ}
\end{equation}%
Substituting (\ref{ZZ}) in (\ref{ZZZ}), we finally obtain the formal
relation between the old and new set of macrovariables labelling the $2l$%
-dimensional macrostates $\Theta $ of generalized Gaussian statistical
models in presence of non-trivial off-diagonal terms,%
\begin{equation}
\mu _{k}\left( \tilde{\mu}_{k}\text{, }\tilde{\sigma}_{k}\right) \overset{%
\text{def}}{=}\tilde{\mu}_{k}+\tilde{\sigma}_{k}\text{ and, }\sigma \left( 
\tilde{\mu}_{k}\text{, }\tilde{\sigma}_{k}\right) \overset{\text{def}}{=}%
\frac{1-\sqrt{\Delta \left( r_{k}\right) }}{2r_{k}}\tilde{\mu}_{k}+\frac{1+%
\sqrt{\Delta \left( r_{k}\right) }}{2r_{k}}\text{ }\tilde{\sigma}_{k}\text{,
with }k=1\text{,..., }l\text{.}
\end{equation}%
After having introduced the information geometric diagonalization procedure,
the new line element $ds^{\prime 2}\left( \tilde{\mu}_{1}\text{,..., }\tilde{%
\mu}_{l}\text{; }\tilde{\sigma}_{1}\text{,..., }\tilde{\sigma}_{l}\right) $
to be considered becomes,%
\begin{equation}
ds^{\prime 2}\left( \tilde{\mu}_{1}\text{,..., }\tilde{\mu}_{l}\text{; }%
\tilde{\sigma}_{1}\text{,..., }\tilde{\sigma}_{l}\right) =\sum_{k=1}^{l}%
\left[ \frac{\alpha _{-}\left( r_{k}\right) }{\left[ a_{0}\left(
r_{k}\right) \tilde{\mu}_{k}+a_{1}\left( r_{k}\right) \tilde{\sigma}_{k}%
\right] ^{2}}d\tilde{\mu}_{k}^{2}+\frac{\alpha _{+}\left( r_{k}\right) }{%
\left[ a_{0}\left( r_{k}\right) \tilde{\mu}_{k}+a_{1}\left( r_{k}\right) 
\tilde{\sigma}_{k}\right] ^{2}}d\tilde{\sigma}_{k}^{2}\right] \text{,}
\end{equation}%
where,%
\begin{equation}
\alpha _{\pm }\left( r_{k}\right) \overset{\text{def}}{=}\frac{3\pm \sqrt{%
\Delta \left( r_{k}\right) }}{2}\text{, }a_{0}\left( r_{k}\right) \overset{%
\text{def}}{=}\frac{1-\sqrt{\Delta \left( r_{k}\right) }}{2r_{k}}\text{, }%
a_{1}\left( r_{k}\right) \overset{\text{def}}{=}\frac{1+\sqrt{\Delta \left(
r_{k}\right) }}{2r_{k}}\text{ and, }\Delta \left( r_{k}\right) \overset{%
\text{def}}{=}1+4r_{k}^{2}\text{.}
\end{equation}%
At this point, omitting additional technical details \textbf{(}that may be
reproduced by following the work presented in Section IV\textbf{)}, we are
able to compute the asymptotic temporal behavior of the dynamical complexity
of geodesic trajectories for the correlated $2l$-dimensional Gaussian
statistical model. It turns out that,%
\begin{equation}
\mathcal{S}_{\mathcal{M}_{s}}\left( \tau \text{; }\left\{ \lambda
_{k}\right\} \text{, }\left\{ r_{k}\right\} \right) \overset{\tau
\rightarrow \infty }{\sim }\sum_{k=1}^{l}\log \left[ \Lambda _{1}\left(
r_{k}\right) +\frac{\Lambda _{2}\left( r_{k}\text{, }\lambda _{k}\right) }{%
\tau }\right]
\end{equation}%
where,%
\begin{equation}
\Lambda _{1}\left( r_{k}\right) \overset{\text{def}}{=}\frac{2r_{k}\sqrt{%
2-r_{k}^{2}}}{1+\sqrt{1+4r_{k}^{2}}}\text{, }\Lambda _{2}\left( r_{k}\text{, 
}\lambda _{k}\right) \overset{\text{def}}{=}\frac{\sqrt{\left(
1+4r_{k}^{2}\right) \left( 2-r_{k}^{2}\right) }}{r_{k}}\frac{\ln \Sigma
\left( r_{k}\text{, }\lambda _{k}\text{, }\alpha _{\pm }\right) }{\lambda
_{k}}\text{, }\alpha _{\pm }\left( r_{k}\right) \overset{\text{def}}{=}\frac{%
3\pm \sqrt{1+4r_{k}^{2}}}{2}\text{. }
\end{equation}%
The quantity $\Sigma \left( r_{k}\text{, }\lambda _{k}\text{, }\alpha _{\pm
}\right) $ is a strictly positive function of its arguments. For $%
r_{k}=r_{s} $, $\forall k$, $s=1$,..., $l$, the information geometric
entropy $\mathcal{S}_{\mathcal{M}_{s}}\left( \tau \text{; }l\text{, }\lambda 
\text{, }r\right) $ becomes,%
\begin{equation}
\mathcal{S}_{\mathcal{M}_{s}}\left( \tau \text{; }l\text{, }\lambda \text{, }%
r\right) \overset{\tau \rightarrow \infty }{\sim }\log \left[ \Lambda
_{1}\left( r\right) +\frac{\Lambda _{2}\left( r\text{, }\lambda \right) }{%
\tau }\right] ^{l}\text{.}
\end{equation}%
Therefore, the information geometric complexity presents a power law decay
where the power is related to the cardinality $l$ of the microscopic degrees
of freedom characterized by correlated pieces of macroscopic information and
it approaches a saturation value quantified by the set of $\left\{
r_{k}\right\} $, the correlation strengths.

\section{Concluding Remarks}

In this article, we have explored the possibility of describing the
macroscopic behavior of complex systems in terms of the underlying
statistical structure of their microscopic degrees of freedom by the use of
statistical inference and information geometry. Specifically,\textbf{\ }we
have computed the asymptotic temporal behavior of the dynamical complexity
associated with the maximum probability trajectories on $2l$\textbf{-}%
dimensional Gaussian statistical manifolds in presence of macro-correlations
between the macrovariables labeling the macrostates of the system. The
algorithmic structure of our asymptotic computations on curved statistical
manifolds is presented in detail\ for the two-dimensional case while it is
schematically outlined for the\textbf{\ }$2l$\textbf{-}dimensional case.
Special focus was devoted to the information geometric diagonalization
procedure that allows to simplify the integration of coupled systems of
nonlinear second order ordinary differential equations\textbf{. }It was
found that the introduction of additional information constraints
(correlations) between the macrovariables of the Gaussian statistical model
led to the emergence of an \textit{asymptotic information geometric
compression} of the explored statistical macrostates on the configuration
manifold of the model in its evolution between the initial and final
macrostates. Actually, the presence of correlations between relevant
macroscopic information about the microscopic degrees of freedom of a
complex system under consideration allows for an information geometric
description whose complexity, measured in terms of\textbf{\ }$\mathcal{S}_{%
\mathcal{M}_{s}}$\textbf{\ (}or\textbf{\ }$\mathcal{V}_{\mathcal{M}_{s}}$%
\textbf{), }decays according to a power law,%
\begin{equation}
\mathcal{S}_{\mathcal{M}_{s}}\left( \tau \text{; }\lambda _{1}\text{,..., }%
\lambda _{l}\text{, }r_{1}\text{,..., }r_{l}\right) \overset{\tau
\rightarrow \infty }{\sim }\sum_{k=1}^{l}\log \left[ \Lambda _{1}\left(
r_{k}\right) +\frac{\Lambda _{2}\left( r_{k}\text{, }\lambda _{k}\right) }{%
\tau }\right] \text{.}
\end{equation}%
Asymptotically, the complexity achieves a saturation value characterized
solely by the strength of such correlations. The relevance of these results
becomes evident when compared to what occurs in absence of correlations\ 
\cite{cafaroIJTP, cafaroPD}, namely%
\begin{equation}
\mathcal{S}_{\mathcal{M}_{s}}\left( \tau \text{; }\lambda _{1}\text{,..., }%
\lambda _{l}\right) \overset{\tau \rightarrow \infty }{\sim }%
\sum_{k=1}^{l}\log \left[ \exp \left( \lambda _{k}\tau \right) \right] \text{%
.}
\end{equation}%
In absence of correlations, the information geometric entropy of a\textbf{\ }%
$2l$\textbf{-}dimensional Gaussian model increases linearly in time and its
complexity diverges exponentially at a rate determined by\textbf{\ }$\left\{
\lambda _{k}\right\} $\textbf{, }the Lyapunov exponents of the statistical
trajectories of the system \cite{carlo-tesi}.

We are confident the information geometric techniques introduced in this
work constitute a further step towards the computation of the asymptotic
behavior of the dynamical complexity of microscopically correlated
multidimensional Gaussian statistical models \cite{ali} and other models of
relevance in more realistic physical systems. Our ultimate hope is to extend
this approach in the field of Quantum Information to better understand the
connection between quantum entanglement and quantum complexity \cite%
{nielsen, cafaroPA, prosen, benenti}.

\begin{acknowledgments}
This work was supported by the European Community's Seventh Framework
Program under grant agreement 213681 (CORNER Project; FP7/2007-2013).
\end{acknowledgments}


\begin{thebibliography}{99}
\bibitem{gell-mann} M. Gell-Mann, "\emph{What is Complexity?}", Complexity,
vol. \textbf{1}, no. 1, 16-19 (1995).

\bibitem{caticha1} A. Caticha, "\emph{Entropic Dynamics}", in \textit{%
Bayesian Inference and Maximum Entropy Methods in Science and Engineering},
ed. by R.L. Fry, AIP Conf. Proc. \textbf{617}, 302 (2002).

\bibitem{carlo-tesi} C. Cafaro, "\emph{The Information Geometry of Chaos}",
Ph. D. Thesis, State University of New York at Albany, Albany, NY, 2008.

\bibitem{carlo-CSF} C. Cafaro, "\emph{Works on an information
geometrodynamical approach to chaos}", Chaos, Solitons \& Fractals \textbf{41%
}, 886 (2009).

\bibitem{caticha2} A. Giffin, "\emph{Maximum Entropy: The Universal Method
for Inference}", Ph. D. Thesis, SUNY at Albany, NY-USA (2008); A. Caticha
and R. Preuss, "\emph{Maximum entropy and Bayesian data analysis: Entropic
prior distributions}", Phys. Rev. \textbf{E70}, 046127 (2004).

\bibitem{amari} S. Amari and H. Nagaoka, \emph{Methods of Information
Geometry}, American Mathematical Society, Oxford University Press, 2000.

\bibitem{catichaIED} A. Caticha, "\emph{Change, Time and Information Geometry%
}", in "Maximum Entropy and Bayesian Methods in Science and Engineering" ed.
by A. Mohammad-Djafari, AIP Conf. Proc. \textbf{568}, 72 (2001).

\bibitem{casetti} L. Casetti, C. Clementi, and M. Pettini, "\emph{Riemannian
theory of Hamiltonian chaos and Lyapunov exponents}", Phys. Rev. \textbf{E54}%
, 5969 (1996).

\bibitem{di bari} M. Di Bari and P. Cipriani, "$\emph{Geometry}$ $\emph{and}$
$\emph{Chaos}$ $\emph{on}$ $\emph{Riemann}$ $\emph{and}$ $\emph{Finsler}$ $%
\emph{Manifolds}$", Planet. Space Sci. \textbf{46}, 1543 (1998).

\bibitem{jacobi} C. G. J. Jacobi, "\emph{Vorlesungen uber Dynamik}", Reimer,
Berlin (1866).

\bibitem{kawabe} T. Kawabe, "\emph{Indicator of chaos based on the
Riemannian geometric approach}", Phys. Rev. \textbf{E71}, 017201 (2005); T.
Kawabe, "\emph{Chaos based on Riemannian geometric approach to Abelian-Higgs
dynamical system}", Phys. Rev. \textbf{E67}, 016201 (2003).

\bibitem{tribus} M. Tribus, "\emph{Rational Descriptions,\ Decisions and
Designs}", Pergamon Press (1969).

\bibitem{caticha(REII)} A. Caticha, "\emph{Relative Entropy and Inductive
Inference}", \textit{Bayesian Inference and Maximum Entropy Methods in
Science and Engineering},ed. by G. Erickson and Y. Zhai, AIP Conf. Proc.%
\textbf{\ 707}, 75 (2004).

\bibitem{caticha-giffin} A. Caticha and A. Giffin, "\emph{Updating
Probabilities}", in \textit{Bayesian Inference and Maximum Entropy Methods
in Science and Engineering, ed.} by Ali Mohammad-Djafari, AIP Conf. Proc. 
\textbf{872}, 31 (2006).

\bibitem{newJay} E. T. Jaynes, "\emph{Information theory and statistical
mechanics, I}", Phys. Rev. \textbf{106}, 620 (1957); \ "\emph{Information
theory and statistical mechanics, II}", Phys. Rev. \textbf{108}, 171 (1957).

\bibitem{cafaroIJTP} C. Cafaro, \textquotedblleft \emph{%
Information-Geometric Indicators of Chaos in Gaussian Models on Statistical
Manifolds of Negative Ricci Curvature}\textquotedblright , Int. J. Theor.
Phys. \textbf{47}, 2924 (2008).

\bibitem{felice} F. De Felice and J. S. Clarke, "\emph{Relativity on Curved
Manifolds}", Cambridge University Press (1990).

\bibitem{cafaroPD} C. Cafaro and S. A. Ali, "\emph{Jacobi Fields on
Statistical Manifolds of Negative Curvature}", Physica \textbf{D234}, 70
(2007).

\bibitem{ali} S. A. Ali, C. Cafaro, D.-H. Kim, S. Mancini, "\emph{The Effect
Of Microscopic Correlations On The Information Geometric Complexity Of
Gaussian Statistical Models}", Physica \textbf{A389}, 3117 (2010).

\bibitem{nielsen} M. A. Nielsen, "\emph{Quantum information science as an
approach to complex quantum systems}", arXiv:quant-ph/0208078 (2002).

\bibitem{cafaroPA} C. Cafaro and S. A. Ali, "\emph{Can chaotic quantum
energy levels statistics be characterized using information geometry and
inference methods?}", Physica \textbf{A387}, 6876 (2008).

\bibitem{prosen} T. Prosen, "\emph{Chaos and Complexity of Quantum Motion}",
J. Phys. \textbf{A40}, 7881 (2007).

\bibitem{benenti} G. Benenti and G. Casati, "\emph{How complex is quantum
motion?}", Phys. Rev \textbf{E79}, 025201 (2009).
\end{thebibliography}
\end{document}